\documentclass{iau} 

\usepackage{graphicx}

\usepackage{hyperref} 

\def\aj{AJ}%
\def\apj{ApJ}%
\def\apjs{ApJS}%
\def\aaps{A\&AS}%
\def\mnras{MNRAS}%
\def\pasp{PASP}%
\def\ssr{Space~Sci.~Rev.}%
\newcommand{\citep}[1]{\cite{#1}}

\title[The Hubble Catalog of Variables] 
{The Hubble Catalog of Variables (HCV)}

\author[Sokolovsky, et al.]   
{K.~V.~Sokolovsky$^{1,2,3}$,
A.~Z.~Bonanos$^{1}$,
P.~Gavras$^{1}$,
M.~Yang$^{1}$,
D.~Hatzidimitriou$^{4,1}$,
M.~I.~Moretti$^{5,1}$,
A.~Karampelas$^{6,1}$,
I.~Bellas-Velidis$^{1}$,
Z.~Spetsieri$^{1,4}$,
E.~Pouliasis$^{1,4}$,
I.~Georgantopoulos$^{1}$,
V.~Charmandaris$^{1}$,
K.~Tsinganos$^{1}$,
N.~Laskaris$^{7}$,
G.~Kakaletris$^{7}$,
A.~Nota$^{8,9}$,
D.~Lennon$^{10}$,
C.~Arviset$^{10}$,
B.~C.~Whitmore$^{8}$,
T.~Budavari$^{11}$,
R.~Downes$^{8}$,
S.~Lubow$^{8}$,
A.~Rest$^{8}$,
L.~Strolger$^{8}$,
R.~White$^{8}$}

\affiliation{$^{1}$IAASARS, National Observatory of Athens, 15236 Penteli, Greece \\[\affilskip]
$^{2}$Sternberg Astronomical Inst. MSU, Universitetskii~pr. 13, 119992 Moscow, Russia \\[\affilskip]
$^{3}$Astro Space Center, LPI RAS, Profsoyuznaya Str. 84/32, 117997 Moscow, Russia \\[\affilskip]
$^{4}$Department of Physics, National and Kapodistrian University of Athens, 15771 Ilissia, Greece \\[\affilskip]
$^{5}$INAF-Osservatorio Astronomico di Capodimonte, Salita Moiariello, 16, 80131 Napoli, Italy \\[\affilskip]
$^{6}$American Community Schools of Athens, 129 Aghias Paraskevis Ave., 15234 Halandri, Greece \\[\affilskip]
$^{7}$Athena Research and Innovation Center, Artemidos 6 \& Epidavrou, 15125~Maroussi, Greece \\[\affilskip]
$^{8}$Space Telescope Science Institute, 3700 San Martin Drive, Baltimore, MD 21218, USA \\[\affilskip]
$^{9}$European Space Agency, Research and Scientific Support Dep., Baltimore, MD 21218, USA \\[\affilskip]
$^{10}$European Space Astronomy Centre, PO\,78, Villanueva de la Ca\~{n}ada, 28692 Madrid, Spain \\[\affilskip]
$^{11}$The Johns Hopkins University, Baltimore, MD 21218, USA}

\pubyear{2018}
\volume{339}  
\jname{Southern Horizons in Time-Domain Astronomy}
\editors{E. Griffin, R. Seaman, M. Sullivan \& P. Woudt eds.}
\begin{document}

\maketitle

\begin{abstract}
The Hubble Source Catalog (HSC) combines lists of sources detected on images 
obtained with the WFPC2, ACS and WFC3 instruments aboard the Hubble Space Telescope (HST)
available in the Hubble Legacy Archive. The catalog contains time-domain
information with about two million of its sources detected with the same instrument and filter 
in at least five HST visits.
The Hubble Catalog of Variables (HCV) project aims to identify HSC sources showing
significant brightness variations.
A magnitude-dependent threshold in the median absolute deviation of photometric
measurements (an outlier-resistant measure of lightcurve scatter) is adopted
as the variability-detection statistic. It is supplemented with a cut in $\chi_{\rm red}^2$
that removes sources with large photometric errors. A pre-processing
procedure involving bad image identification, outlier rejection and
computation of local magnitude zero-point corrections is applied to 
HSC lightcurves before computing the variability detection statistic.
About 52\,000 HSC sources are identified as candidate variables, 
among which 7\,800 show variability in more than one filter.
Visual inspection suggests that $\sim 70\%$ of the candidates detected in
multiple filters are true variables while the remaining $\sim 30\%$ are sources
with aperture photometry corrupted by blending, imaging artifacts or image 
processing anomalies. The candidate variables have AB magnitudes in the range 15--27$^{m}$ with the median 22$^{m}$.
Among them are the stars in our own and nearby galaxies as well as active galactic nuclei.

\keywords{techniques: photometric, stars: variables: other}
\end{abstract}

\firstsection 

\section{Introduction}

The Hubble Source Catalog (HSC; \cite{2016AJ....151..134W,2012ApJ...761..188B}) combines
individual source lists derived from images obtained with the WFPC2, 
ACS and WFC3 cameras aboard the Hubble Space Telescope (HST).
These source lists are created with SExtractor (\cite{1996A&AS..117..393B})
from stacked images combining (with AstroDrizzle; \cite{2012AAS...22013515H}) 
individual exposures taken within one HST visit. 
Image stacking and source list extraction are performed as part of 
the routine processing in the Hubble Legacy Archive
(\cite{2006ASPC..351..406J,2008ASPC..394..481W}).
As the accuracy of astrometric solutions associated with original HST images 
is limited by the positional accuracy of individual Guide Star Catalog
(\cite{2008AJ....136..735L}) stars, the HSC is cross-matched with the deep
catalogs of Pan-STARRS, 2MASS, and SDSS to reach a typical absolute astrometric 
accuracy of $<0.1^{\prime\prime}$. The Gaia catalog will be used to further increase the accuracy of absolute astrometry in future HSC releases. The HSC sources are distributed in isolated pencil beams covering about 0.1\% of the sky.
For each detected source the HSC reports aperture photometry results 
based on published instrument zero-points as described at 
\url{http://hla.stsci.edu/hla_faq.html}
The HSC photometric precision estimated from repeated measurements of the
same sources is about 0.02--0.04$^{m}$.

One of the primary science goals of the HST is to improve the
extragalactic distance scale (\cite{2018SSRv..214...32C}) by observing standard candles 
including Cepheids (\cite{2001ApJ...553...47F}), supernovae (\cite{2018ApJ...853..126R}) and Mira variables
\cite{2018arXiv180102711H}.
Some fields were observed in multiple HST visits to obtain very deep mosaic images
(\cite{2013ApJS..209....3K}),
perform astrometric studies (e.g. \cite{2010ApJ...710.1032A,2014MNRAS.442.2381N}) or calibration
(\cite{2011PASP..123..622B}).
The Hubble Catalog of Variables (HCV) project aims to use the time domain
information from the HSC to perform a uniform variability search in the fields
visited by the HST multiple times.

In Section~\ref{sec:pre} we summarize our variability detection technique
and present preliminary variability search results in Section~\ref{sec:res}
Previous reports on the state of the HCV project were presented by 
\cite{2017IAUS..325..369G}, \cite{2017EPJWC.15202005S}, \cite{2017arXiv171111491Y}.

\section{Pre-processing and variability detection}
\label{sec:pre}

The HSC data are naturally grouped in clusters of sources detected on
spatially overlapping images. For all sources in such a group we extract
lightcurves in all instrument/filter combinations with which this particular
area of the sky was observed. A set of lightcurves in one group 
obtained with the same instrument and filter is our basic variability search
unit. The data quality criteria and variability detection thresholds are
determined for each such unit independently.

Prior to conducting a variability search we try to improve the quality of
the input photometric data. 
The first step is to apply quality cuts to discard sources marked as saturated and detected less than a specified number of times (currently 5) with the given instrument and filter (so the lightcurves of all the considered sources have at least the specified number of points).
We reject groups where fewer than 300 sources pass the quality cuts, in all instrument-filter combinations, as smaller samples are less well suited for the statistical analysis followed.

We assign a weight to each lightcurve point using the ``synthetic error'' -- 
a combination of the estimated photometric error, the magnitude difference between 
the two concentric apertures of a different diameter (``concentration
index''), 
the magnitude difference between the circular and automatically selected elliptical
aperture tuned to the source size (SExtractor parameter ``MAG\_AUTO'') 
and the offset distance from the matching position. Elevated values of any of
these parameters may indicate that the source is blended, affected by an
uncleaned cosmic ray or other imaging artifact. For each lightcurve we
perform a weighted robust linear fitting. Lightcurve points deviating from
this fit or the ones with high values of the synthetic error are flagged as
outliers. Visits resulting in a high percentage of outliers ($>20\%$) are
identified as bad and all measurements corresponding to such visits are
discarded. 

The second pre-processing step is the calculation and application
of local magnitude zero-point corrections (e.g. \cite{2014MNRAS.442.2381N})
that minimize the impact of large-scale sensitivity variations across the
instrument's field of view. For each source we use all other sources within 
a $20^{\prime\prime}$ radius to determine the local correction value for
this source and visit. The local correction is the median difference between
the magnitude predicated by the robust linear fit to the lightcurve and the
actual measured magnitude.
Variability detection in the lightcurves
obtained with the same instrument and filter require maximizing the 
relative precision while the absolute photometric accuracy is of little
concern.

\begin{figure*}[!htb]
 \centering
 \includegraphics[width=0.65\textwidth]{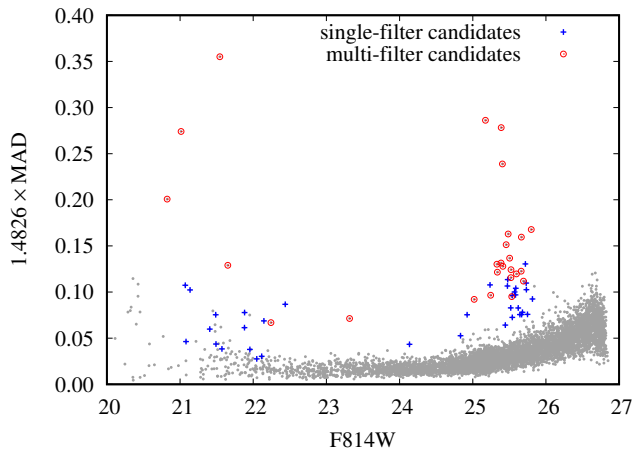}
 \caption{Median absolute deviation (scaled to $\sigma$ of the Gaussian distribution) as a function of F814W magnitude for a field in the halo of M\,31 originally investigated by \cite{2004AJ....127.2738B}. The plot highlights candidate variables selected using F814W or F606W data (single-filter) and the ones found in both filters (multi-filter candidates).}
 \label{fig:mad}
\end{figure*}

We searched for a variability detection method that would be sensitive to 
a wide range of variability types, robust to outlier measurements and
applicable to lightcurves with a small number of points (the last requirement
rules out the period search techniques).
After comparing 18 ``variability indices'' discussed by \cite{2017MNRAS.464..274S}
we selected a magnitude-dependent cut in Median Absolute Deviation (MAD) as our 
variability detection statistic.
MAD is a robust measure of lightcurve scatter (e.g. \cite{2016PASP..128c5001Z}) defined as 
$${\rm MAD} = {\rm median}(|m_i-{\rm median}(m_i)|)$$ 
where $m_i$ is the $i$th magnitude measurement in a lightcurve.
For each group and each set of measurements obtained with the same instrument
and filter we identify as candidate variables the lightcurves having their
MAD values at least $5\sigma$ larger than the median MAD value at the source (median) magnitude. 
Here we rely on the assumption that the majority of
field sources are not variable above the 1\% level, so the ones that stand out
in the MAD--magnitude plot (Fig.~\ref{fig:mad}) are the variable sources or sources measured with
much lower accuracy than the majority of sources in this group.
Photometric errors reported in the HSC are often underestimated
(\cite{2016AJ....151..134W}), but they are not expected to be overestimated.
We require all candidate variables to have the reduced $\chi^2$ value associated with the hypothesis that the candidate has constant brightness $\chi_{\rm red}^2>3$
(\cite{2010arXiv1012.3754A}). This helps to handle the cases where a
particular object could not be measured well due to high local background
and this was correctly reflected in the errorbars.

\section{Preliminary results}
\label{sec:res}

In our preliminary analysis based on the second version of the HSC, 
out of 1.85 million sources that pass the initial quality cuts for the
variability search, 52\,000 (2.8\%) are marked as candidate variables,
among which 7\,800 are identified as variable in more than one filter.
The candidates have the AB magnitude range of 15--27$^{m}$ (median 22$^{m}$).

We performed visual inspection of all the candidate variables detected in
multiple filters. For each candidate we inspected its lightcurve, position
on the color-magnitude diagram and the three images associated with 
the brightest, median and faintest brightness measurement. 
About 70\% of the candidates pass the visual inspection having no obvious
problems with their images (crowding, imaging artifacts) or similarities
in their lightcurves to the lightcurves of other sources in the group
(that may indicate systematics affecting photometry of multiple
sources).
We noticed that some groups have unrealistically high numbers of candidates 
(showing large scatter in their lightcurves). The problem was
traced down to a slight misalignment between the white-light images used to
detect sources and to place the measurement apertures and the filter-combined
images used to obtain measurements in a given filter. This misalignment
will be eliminated in future HSC versions while for the initial version
of the HCV (based on the second version of the HSC) the fields severely
affected by misalignment had to be excluded from the analysis.

We continue to improve our candidate selection and validation criteria. 
The initial release of the HCV will include the clean sample of variables 
detected in multiple filters and validated by human experts as well 
as the extended list of candidate variables detected with the automated selection only.
The HCV will be released later this year. This work is supported by ESA under contract No. 4000112940.

\end{document}